\documentclass[runningheads]{llncs}
\usepackage[utf8]{inputenc}

\usepackage[T1]{fontenc}
\usepackage{amsmath}
\usepackage{graphicx}
\begin{document}
\title{A User-Centric, Privacy-Preserving, and Verifiable Ecosystem for Personal Data Management and Utilization}
\titlerunning{Privacy-Preserving Ecosystem for Personal Data Management}
%
\author{Osama Zafar\inst{1}\orcidID{0009-0008-9621-6899} \and
Mina Namazi\inst{1}\orcidID{0000-0002-8878-9362} \and
Yuqiao Xu\inst{1}\orcidID{0009-0009-3552-2136} \and
Youngjin Yoo\inst{2}\orcidID{0000-0001-8548-3475} \and
Erman Ayday\inst{1}\orcidID{0000-0003-3383-1081}}   
\authorrunning{Osama. et al.}
%

\institute{Case Western Reserve University, Cleveland, OH 44106, USA
\email{\{oxz23,mxn559,yxx914,exa208\}@case.edu}\\
\url{http://www.springer.com/gp/computer-science/lncs} \and
The London School of Economics and Political Science (LSE)\\
\email{y.yoo@lse.ac.uk}}
\maketitle              
\begin{abstract}
In the current paradigm of digital personalized services, the centralized management of personal data raises significant privacy concerns, security vulnerabilities, and diminished individual autonomy over sensitive information. Despite their efficiency, traditional centralized architectures frequently fail to satisfy rigorous privacy requirements and expose users to data breaches and unauthorized access risks. This pressing challenge calls for a fundamental paradigm shift in methodologies for collecting, storing, and utilizing personal data across diverse sectors, including education, healthcare, and finance.

This paper introduces a novel decentralized, privacy-preserving architecture that handles heterogeneous personal information, ranging from educational credentials to health records and financial data. Unlike traditional models, our system grants users complete data ownership and control, allowing them to selectively share information without compromising privacy. The architecture's foundation comprises advanced privacy-enhancing technologies, including secure enclaves and federated learning, enabling secure computation, verification, and data sharing. The system supports diverse functionalities, including local computation, model training, and privacy-preserving data sharing, while ensuring data credibility and robust user privacy.

\vspace{-2 mm}

\keywords{Privacy-Enhancing Technologies \and Decentralized Data Management \and Verifiable Computation.}
\end{abstract}

\footnotetext[1]{The paper will be appearing in ESORICS2025.}

\vspace{-7 mm}
\section{Introduction}
\vspace{-2 mm}

Managing personal data has become a defining challenge in modern digital systems. As organizations increasingly collect and store vast amounts of sensitive information in cloud-based centralized architectures, a fundamental conflict has emerged between service efficiency and individual privacy rights. They encounter growing concerns about privacy, security vulnerabilities, and maintaining personal control over data. While efficient, these centralized systems introduce significant vulnerabilities, such as a single point of failure and a lack of user control over sensitive information; thereby demanding a comprehensive reassessment of data governance and personal data management frameworks. 

Traditional protocols often fail to meet robust privacy requirements, exposing users to data breaches and unauthorized access. This issue is particularly evident in various sectors. Data centralization has transformed these systems into prime targets for cybersecurity attacks, prompting users to be concerned about the security and sovereignty of their data.

Such problems manifest particularly in sectors handling extremely sensitive personal information. Healthcare organizations manage extensive collections of electronic health records and data from smart devices. Educational institutions handle student records and learning analytics. Financial institutions deal with confidential transactions and personal financial information. The rise of IoT devices and increased connectivity has further accelerated the exponential growth of personal data, amplifying the need for stricter privacy regulations and innovative approaches to data management.

Existing centralized platforms face three fundamental limitations. First, they create security vulnerabilities through data concentration, making them targets for cyberattacks~\cite{aslan2023}. Second, they disenfranchise individuals from data ownership and portability, allowing service providers to manage personal data without adequate user control~\cite{battiston2023}. They obtain consent from users to use their data; however, they collect raw user data even when it is not needed, leading to a loss of control and ownership over the data. Third, despite their centralized nature, these systems often fail to achieve seamless data integration, resulting in incomplete or fragmented data collections.  
On the other hand, current decentralized solutions address some privacy concerns~\cite{alazab2024privacy,bernal2019privacy}, however significant gaps remain in developing comprehensive functionalities required for modern digital services. Current solutions often struggle to perform secure computations on private data or facilitate controlled sharing between service providers. Furthermore, existing systems struggle to maintain data utility while preserving privacy, particularly in scenarios requiring complex data analysis or machine learning applications \cite{mireshghallah2021privacy,synergizing2024privacy}.

To address these limitations, in this work, we propose a privacy-preserving, decentralized, AI-enabled data ecosystem.
At the core of our architecture are user-controlled decentralized entities called 'data agents' that serve as secure vaults for personal data storage and computation. These data agents provide users with privacy by design and return control while enabling complex data utilization. Similar to a traditional framework, the proposed architecture utilizes a consent mechanism in the form of access control management (allowing users to configure service providers' access levels to their data) while keeping the raw data within the user's data agent.

Our architecture incorporates several key technical innovations. First, it implements privacy-preserving computation capabilities, allowing service providers to run analyses without accessing underlying raw data. By integrating federated learning methodologies, our system enables collaborative model training across distributed data sources without centralizing sensitive information. Additionally, it provides domain-specific customization options to address the unique requirements of different sectors while maintaining consistent privacy guarantees. The use of secure enclaves allows the system to execute computations in a secure and verifiable manner by generating cryptographic signatures with the results as proof against tampering. Building on these capabilities, our architecture delivers privacy-preserving analytics by responsibly managing and leveraging personal data. It meets the dual demands of privacy and functionality, fostering trust and innovation in an increasingly data-driven world. This positions our system as a robust and future-ready platform for addressing the complex needs of modern industry applications. 

From a technical implementation perspective, the system is platform-agnostic and can be easily deployed to different cloud computing platforms. We utilize AWS services as a case study. The proposed system leverages AWS Nitro enclaves \cite{AWSNitroEnclaves} for secure computation and verifiability, ensuring that even when processing user data, proprietary models and algorithms from service providers remain confidential. Our architecture includes novel data plugs that enable secure data collection from various sources while maintaining user control through comprehensive access management tools. Integrating DIDComm \cite{DIDComm} for secure communication ensures end-to-end privacy in all data exchanges. Robust security is a cornerstone of our architecture, implemented through a multilayered approach. At its foundation, data agents implement strict access controls and secure storage using AWS security features, including multi-factor authentication. All data collection includes digital signatures for a verifiable chain of possession. 

A distinctive advantage of our proposed architecture is its ability to consolidate data from various aspects of a user's life, including finance, healthcare, education, social media, entertainment activity, GPS, driving, history, and more, all while retaining ownership with the user. This holistic data integration from the user's entire life allows service providers to develop a more comprehensive profile of the user's personality and deliver enhanced personalized services without compromising the privacy of the users. 
For instance, content platforms like Netflix collect data on users' preferences and recommend content to watch. However, it operates within restricted visibility of user preferences due to limited user activity on the platform. This limits their recommendation capabilities to activities within their specific service.
In contrast, services like YouTube and Spotify collect similar data to recommend content, benefiting from a higher activity level due to their shorter content duration.  Our proposed system enables sophisticated personalization by allowing service providers to analyze patterns across platforms and contexts, all while preserving privacy and user control. In this instance, Netflix can leverage consolidated user activity data to enhance its recommendation algorithms significantly. Similarly, other service providers can deliver highly personalized content that resonates with individual tastes by analyzing user interactions across different platforms, including social media activity and cross-platform engagement.

The proposed solution protects personal and sensitive information in an open ecosystem, providing a multidirectional data flow that allows individuals and small entities to co-create meaningful value from their data. Furthermore, it presents an economic incentive for organizations to maintain and update decentralized datasets, making the overall open data ecosystem more sustainable.

Our system has been designed to address the unique challenges of key domains such as education, healthcare, and finance, making it highly practical and relevant to these critical sectors. The architecture's potential impact extends beyond immediate applications, offering a foundation for future privacy-preserving digital services across diverse industries.

We summarize our main \textbf{contributions} as follows.

\begin{itemize}
    \item Integration of secure and privacy-preserving computation within a decentralized framework, enabling data processing without centralization;
\item Support for federated learning across distributed data agents while maintaining privacy;
\item Comprehensive data agent model that combines secure storage, computation, and sharing capabilities;
\item Practical prototype implementation focused on specific applications such as healthcare, education, and the finance industry requirements.
\end{itemize}

The rest of the paper is organized as follows. We summarize the related work in Section~\ref{related}. The proposed architecture is introduced in Section~\ref{sec:proposed}, and its security and privacy analyses are discussed in Section~\ref{sec:security}. We evaluate the feasibility of our proposed scheme in Section~\ref{sec:evaluation} and discuss its application in Section~\ref{sec:application}. Finally, we discuss extending our framework to include more functionalities and support more robust security definitions in Section~\ref{sec:discussion}, and conclude our research in Section~\ref{sec:conclusion}.

\label{sec:intro}

\vspace{-2 mm}
\section{Related Work}
\label{related}
\vspace{-2 mm}

Public awareness of data rights and privacy has grown significantly in recent years. Regulations such as the General Data Protection Regulation (GDPR) \cite{wikipedia2021gdpr} in the European Union and the California Consumer Privacy Act (CCPA) in the United States underscore the global shift toward improved data protection standards. These legal frameworks impose stricter data management protocols and place the responsibility on organizations to ensure robust consumer data privacy.

In response to these evolving requirements, several platforms have emerged to address the need for privacy-preserving data management. Digi.me \cite{digime} offers a personal data platform that allows users to aggregate and control their data from various sources. While providing users a centralized view of their data, it lacks the advanced computational capabilities and privacy-preserving features such as decentralized model training and secure computation environments. 

MIT's Solid (Social Linked Data) project \cite{Solid} presents a decentralized data storage system, giving users control over where their data is stored and managed. It enables users to store their data in personal online data stores called "Pods." Users have the freedom to choose where these Pods are hosted and can grant permissions to specific people to access parts of their data. Similarly, OpenPDS (Personal Data Store) \cite{openpds}, also developed at the Massachusetts Institute of Technology (MIT), enables individuals to collect, store, and provide fine-grained access to their data. Both systems offer solution to store and manage personal data with full access control. However, they do not guarantee complete data ownership, as shared data with third parties become part of their centralized system and can not easily be recalled.

Dataswift's PDA (Personal Data Account) \cite{dataswift} provides an infrastructure for individual data ownership and control. It enable individuals to collect, store, process and use their own personal data in the cloud. It allows users to permit third party applications to read and write data. However, it lacks control to the subsequent storage, processing, and use of user's data by the authorized third parties.


PersonalData.IO~\cite{personaldataio} is a platform designed to empower individuals by providing them with tools and resources to control their personal data. It focuses on creating transparency and accountability in how companies collect, store, and use personal data. By leveraging GDPR compliance and other privacy regulations, PersonalData.IO allows users to understand what personal information companies hold about them, request data access or deletion, and maintain their privacy rights.

Meeco~\cite{meeco} offers a user-centric data management platform that bridges the gap between personal data ownership and ethical data usage. Focusing on secure data sharing and privacy, Meeco bridges the gap between personal data ownership and ethical data usage. Its user-centric approach ensures that individuals can actively manage who has access to their information while maintaining transparency and accountability in how their data is used.

The Databox architecture~\cite{databox} offers a privacy-centric approach through a local data collection and processing system that empowers individuals to control their data while enabling secure third-party sharing. Databox focuses on the local processing of IoT data and eliminates the need to send sensitive information to third party services. It shifts data control from centralized cloud providers to users through a hybrid system with a local physical device and cloud-hosted services that work together to manage personal data collection, storage, and processing. While being a compelling privacy-centric framework for personal data management, it has certain limitations. It primarily focus on IoT data collection and processing, which restricts its ability to aggregate and leverage multi-modal data from diverse third-party services. Although the Databox allows the execution of third-party applications in isolated environments, it lack advanced computational and model training capabilities required by complex service models like recommendation systems. It also lack robust verifiability guarantees against data tampering for the results generated by third-party applications.

Data Bank model \cite{databank} is a privacy-preserving architecture for cloud-IoT platforms designed to protect users' sensitive data by giving them control over what data their devices transmit and providing tools to manage privacy-utility tradeoffs. It incorporates a category-based data access (CBDA) model for managing privacy policies, allowing data owners to define access permissions based on the data category. 

P-PDS (Privacy-Aware Personal Data Storage) \cite{ppds} is a user-centric system designed to automate privacy decisions for third-party access requests based on user preferences. PDS specifically offers individuals the capability to keep their data in unique logical data stores that can be connected and used by proper analytical tools or shared with third parties under the control of end users.

While existing solutions have made important advances in privacy-preserving data management, focusing on decentralized data access, ownership, and governance, they exhibit significant limitations. Current platforms generally fall into one of three categories: (i) solutions that focus primarily on decentralizing data access and providing users ownership over their data (e.g., Digi.me and MIT's Solid), (ii) platforms that emphasize control over data sharing mechanisms (e.g., OpenPDS and Meeco) and (iii) architectures that concentrate on local storage and processing capabilities (e.g., Data Bank and Databox). However, these approaches are constrained by their limited computational capabilities, which impair their ability to perform complex data analysis and extract utility from the data while maintaining privacy. Our proposed architecture distinguishes itself by enabling decentralized data control while seamlessly integrating federated learning and secure computation capabilities without centralizing or exposing personal user data.

This innovative approach allows our architecture to maintain robust privacy guarantees while enabling third-party services to analyze data without exposing sensitive personal information. Unlike localized systems that are restricted due to the limited processing capabilities of local hardware, the proposed architecture leverages federated learning in a cloud computing environment to perform scalable and efficient data analysis, overcoming the computational bottlenecks seen in existing solutions. By allowing secure, privacy-preserving data sharing and analysis,  our system successfully bridges the critical gap between privacy protection and data utility, empowering users while maintaining the value of their data for third-party applications.

Our distinctive design safeguards user privacy and empowers individuals to retain complete ownership of their information while contributing to valuable insights and analytics. Unlike many current systems that depend on integrating user data and transferring raw data to centralized entities, our architecture eliminates the need for such transfers. Users share only the computation results, significantly enhancing privacy and security without compromising the data utility. This approach addresses a significant limitation in existing solutions, which often sacrifice privacy for functionality and fail to deliver actionable insights without risking data exposure.

\vspace{-2 mm}
\section{Proposed Framework}
\label{sec:proposed}
\vspace{-2 mm}

We present our proposed comprehensive, decentralized, privacy-preserving, AI-driven architecture for personal data management. This section details the system model and setting, threat model, and technical components that enable secure and private data processing while maintaining utility for service providers. The used notation is introduced in Table \ref{tab:notation}.

\begin{table}[h]
\centering
\caption{Used Notation}
\label{tab:notation}
\begin{tabular}{|p{3cm}|p{9cm}|}
\hline
\textbf{Notation} & \textbf{Description} \\ \hline
$U= \{u_1, \ldots, u_n\}$ & Set of users in the system \\ \hline
$SP = \{sp_1, \ldots, sp_n\}$ & Set of service providers \\ \hline
$DS = \{d_1, \ldots, d_n\}$ & Set of data sources \\ \hline
$RE = \{re_1, \ldots, re_q\}$ & Set of computation requests \\ \hline
$P$ & Data plug component \\ \hline
$DA$ & Data agent component \\ \hline
$UC$ & User controller component \\ \hline
$MG$ & Model aggregator component \\ \hline
$AC$ & Access control system \\ \hline
$CP$ & Computation policies defining allowed computations\\ \hline
$OP$ & Set of operations \\ \hline
$Auth, ~Cred$ & Authentication system, Credential \\ \hline
$M, ~M^{up}$ & Model, Updated model \\
 \hline
\end{tabular}
\end{table}



\begin{figure}
    \vspace{-5 mm}
    \centering
    \includegraphics[width=0.75\linewidth]{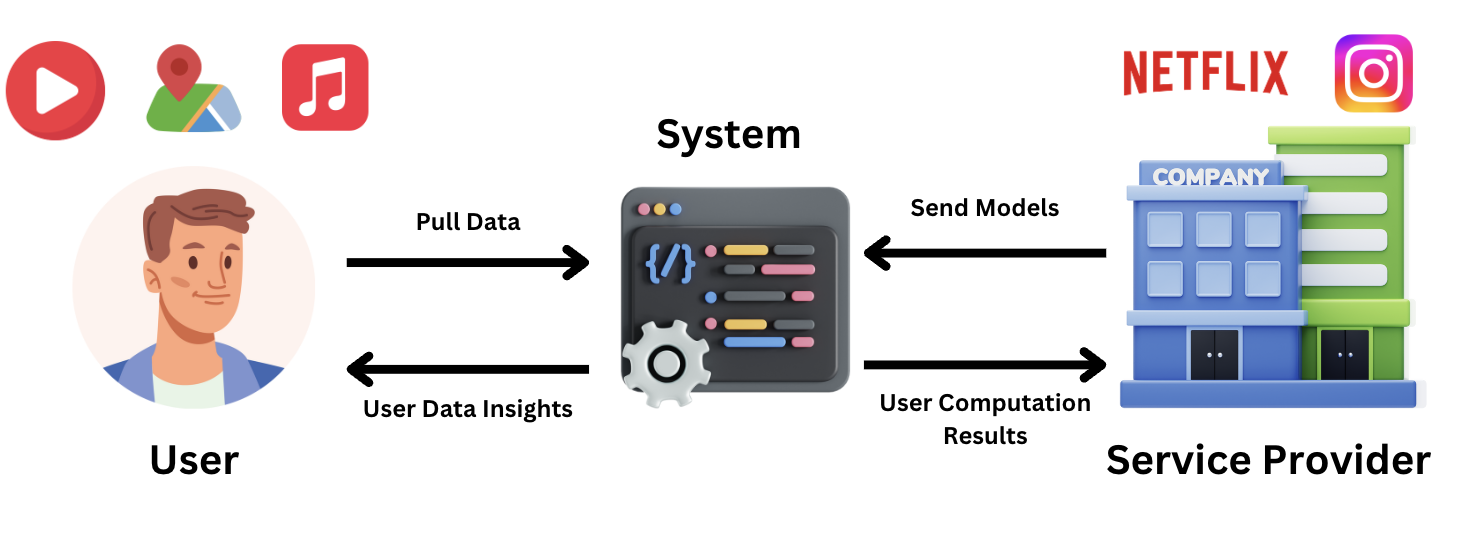}
    \caption{Proposed decentralized, privacy-preserving, AI-driven architecture for personal data management.}
    \label{fig:genflow}
    \vspace{-5 mm}
\end{figure}

\vspace{-2 mm}
\subsection{System Model and Settings}
\label{model}
\vspace{-2 mm}

The proposed architecture comprises two key actors: the set of users $U$ and the set of service providers $SP$. Each party has a component, denoted by \textit{data agent} $DA$, to securely convey, integrate, manage, and aggregate data from different platforms. The user's data agent consists of \textit{data plugs} $(P)$, and \textit{user controllers} $(UC)$. The $SP$'s data agent consists of the service provider controller similar to the $UC$ and a \textit{model aggregator} $(MG)$.

Data plugs $P$ are the data collection components that pull user data through API integration with different raw data sources $DS = \{d_1, d_2, \ldots, d_n \}$ such as medical records, fitness activity records, financial records, and entertainment application activities. $UC$ enforces the access control $AC$ settings for all requests from the $SP$. We implemented a decentralized communication and verification mechanism called DIDComm~\cite{DIDComm} to establish the connection by exchanging decentralized identifiers called DIDs~\cite{DIDs}. The data agent is the central part of the proposed architecture. They are self-identifiers that enable secure communication using verifiable digital identities. $MG$ is a component that the service provider uses to manage its machine learning models and aggregate the users' computation results.

First, data plugs $P$ establish secure connections to various service providers and data sources, pulling data into the system. When a $SP$ intends to utilize this data (for generating personalized recommendations, training machine learning models, or performing analytics), they must submit a request to the user. Every request received is checked by the $UC$ for permission in the $AC$ setting to ensure the user allows the sender (service provider) access to perform a requested action. If the request is approved, $UC$ runs the computation, producing a verifiable result, including a cryptographic attestation $\sigma$ ensuring computational integrity.

Multiple data agents can participate in collaborative scenarios, such as federated learning, while preserving an individual's privacy. Each agent performs local computations (model training) on their data, and the updated trained models are aggregated through secure protocols without exposing raw data. The general framework of the proposed architecture is represented in Figure~\ref{fig:genflow}.

\vspace{-2 mm}
\subsection{Threat Model}
\label{threatmodel}
\vspace{-2 mm}

We define the security of our proposed decentralized, privacy-preserving system against external adversaries and potentially curious or malicious internal parties, including service providers. 

In our framework, the users are the data owners who fully trust their data agents. Service providers are considered honest-but-curious, meaning they follow protocol instructions, but might be curious to learn additional unauthorized information. Data sources are trusted to provide accurate data, but may be compromised. Secure enclaves are trusted for secure computation, and DIDComm is trusted to establish secure communications. External adversaries are considered to have complete network control and can attempt to compromise any participant except the secure enclaves. We acknowledge that securing against side-channel attacks inside secure enclaves is not our concern.

An adversary might attempt to intercept, modify, or inject communications between system components. Our framework prevents these attacks through the DIDComm protocol, which establishes authenticated and encrypted channels between parties using decentralized identifiers (DIDs). Each data source digitally signs its data, creating a verifiable chain of possession. When data agents communicate with service providers or other data agents, they use DIDComm's cryptographic protocols to verify the authenticity of each message.

Malicious actors may manipulate model training results to corrupt the system's output or gain insight into user data through modified computations. Our framework prevents this through a comprehensive verification system. Every computation executed in a secure enclave produces an attestation that cryptographically proves the calculation was performed correctly on legitimate data. For federated learning scenarios, the model aggregator verifies each contribution's attestation before incorporating it into the global model, ensuring that only legitimate, correctly computed updates are included.

Adversaries might attempt to bypass access control mechanisms to gain unauthorized access to user data or computational resources. Each request must satisfy both the permission policy and the computation policy. The system validates all credentials cryptographically and enforces fine-grained permissions through AWS attribute-based access control. Even if an attacker obtains valid credentials, they cannot exceed the explicitly granted permissions, as the user controller component validates each request against the stored access policies before allowing any data access or computation.

The system's decentralized nature, with data stored in individual data agents, eliminates vulnerabilities associated with centralized points of failure. This distributed architecture enhances the system's overall resilience and protects against large-scale data breaches that centralized systems are vulnerable to.

We define the security model of the proposed framework and formally prove it in Section~\ref{sec:security}.

\vspace{-2 mm}
\subsection{Overview}
\label{overview}
\vspace{-2 mm}

Our framework introduces an end-to-end solution that enables service providers to derive valuable insights while allowing users to maintain control over their data. The data plug component aggregates data from external sources (such as Google Maps, Spotify, and YouTube) and stores it in the user's data agent. The user's data agent enables service providers to execute computation functions on the user's data and train machine learning models in a secure and trusted manner. The system utilizes secure enclaves on the user data agents to ensure the verifiability of computations and protect the confidentiality of functions owned by service providers. 



\begin{figure}
    \vspace{-5 mm}
    \centering
    \includegraphics[width =0.75\linewidth]{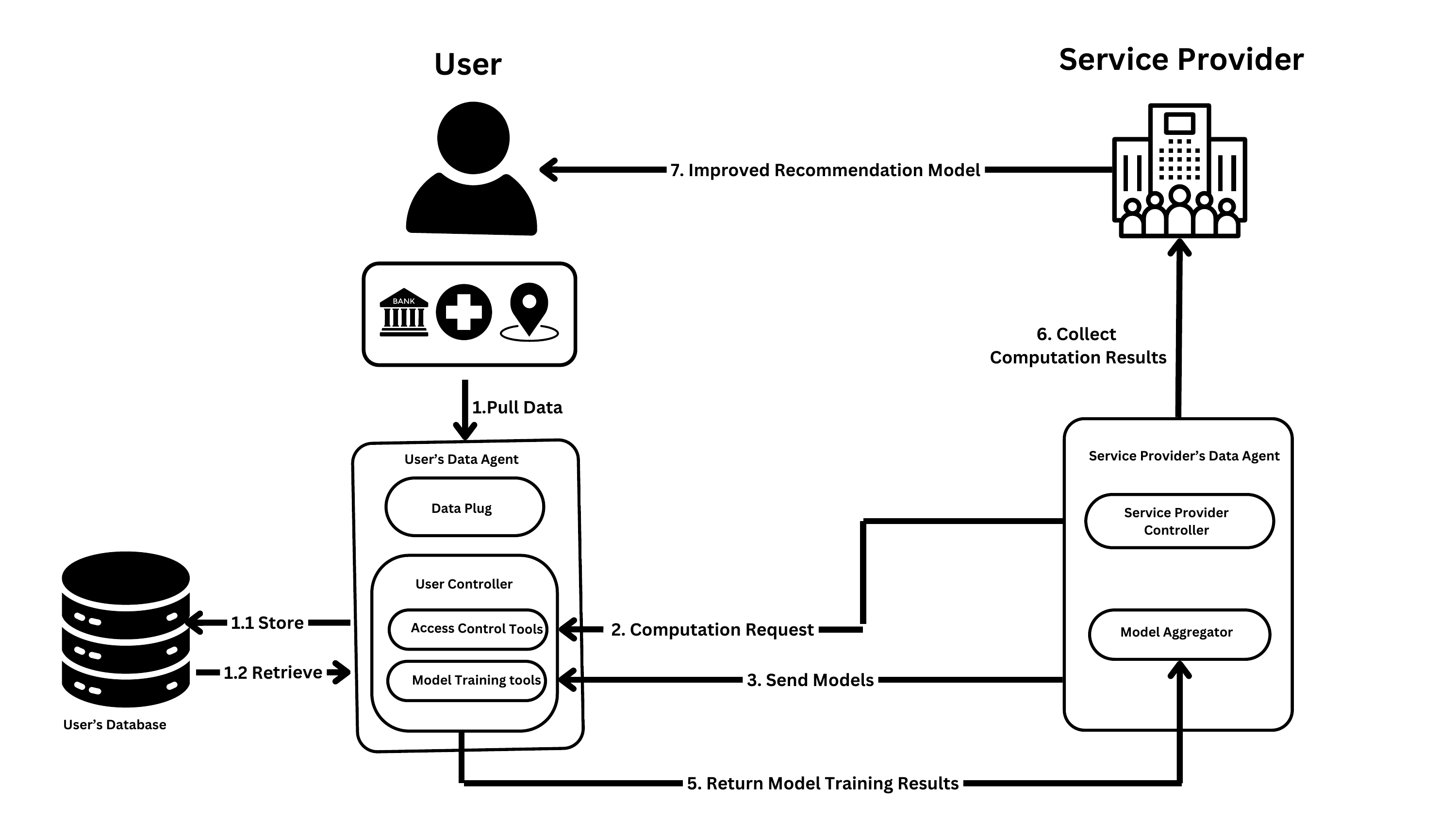}
    \caption{General user flow diagram.}
    \label{fig:userflow}
    \vspace{-7 mm}
\end{figure}

Figure~\ref{fig:userflow} illustrates the general data flow within the proposed architecture, wherein the user retains control over their data through the user's data agent, while the service provider submits requests to perform computations or train machine learning models via the service provider's data agent. 

\vspace{-2 mm}
\subsection{Technical Details}\label{sec:architecture}
\vspace{-2 mm}

The technical details of the proposed framework build upon the formal definitions of data agents and their interactions with service providers through secure protocols and computation mechanisms.

\subsubsection{User's Data Agent Components \\}
\label{userdataagent}

\noindent\textbf{Data plug} component implements a secure data collection and integration mechanism hosted with AWS Elastic Container Service (ECS)~\cite{AWS-ECS}. The data plug component regularly collects the data from third-party sources (such as Google Maps, Spotify, and YouTube) via their APIs and sends it to the user controller component, which pre-processes and stores it in a database for later use. The user can configure and connect the data plug (via an app) to each new data source as it requires i) proof of the user's identity, such as username and password, and ii) the access control settings for the newly added data source. This configuration process for a new data source requires a user's credentials, such as username and password $cred$, and an access control setting for the data source $AC$.
  
\noindent\textbf{User controller} component is the core of the entire architecture. The main functionalities of this component are secure communication and computation. The user controller enforces access control settings for all incoming service provider requests. They check each received request to satisfy the permission in the access control setting, ensuring the user granted the sender (service provider) access to perform a requested action (building a model, computation, and sharing) using the requested piece of the user's data. The user controls the processes of connecting to the data sources and managing them through the access control settings. These settings provide the user with an interface accessible through a web browser or an app to view and manage permission levels for different types of data access. Access control can be configured for each data source, deciding which service provider can access the users' particular data and the operation types they can run. User can update these permissions according to their preferences. To protect stored data, AWS provides a robust security mechanism with two-factor authentication that requires a username, password, and a randomly generated one-time password shared via a previously registered mode of communication like email or phone text message. Furthermore, standardized protocols, such as OpenID Connect (OIDC) and OAuth, can be adopted to provide security beyond usernames and passwords. 

The access control system manages data access permissions for the $U$ through a formal specification
$AC = (U, SP, DS, CP, OP, Auth)$.

The permissions are granted for each data source $DS$ if $Perm(DS) = (SP, OP) \rightarrow 1$.

The access control function evaluates requests $RE$. It allows access if all the computation policies in user-defined access control settings are satisfied, namely if the $Valid$ function outputs $1$ on the inputs, $Valid(RE, CP) \rightarrow 1$, and allows the process, $Allow(RE) = Valid(AC) \land Valid(RE, CP)$.

\subsubsection{Service Provider's Data Agent \\}
\label{seviceprovideragent}

\noindent\textbf{Model aggregator} component, $MG$, is operated by service providers to manage machine learning models and aggregate trained model results from user data agents. When a service provider initiates a model-related request, $MG$ first distributes the computation task to eligible users' data agents. Each user data agent performs local computations (model training) on their private data and returns results with an attestation of the correctness of the result to the $MG$. Then, depending on the setting, $MG$ implements an aggregation protocol and combines individual results. The $MG$ maintains a set of models and manages their updates based on the aggregated results. This process is advantageous for service providers to utilize distributed computations across multiple data agents, while ensuring that individual user data is protected within each data agent. As a result, the raw data of the users is not exposed to the service provider or any other participants.

\noindent\textbf{Service Provider Controller} is similar to the user controller, as it sends requests from the service provider side to perform tasks on the user's data. Hence, we denoted it using the same notion $UC$. 

We deploy a DIDComm Agent~\cite{DIDComm}, which is a communication middleware to establish the decentralized communication connection between parties (between user data agents or between a user data agent and service provider) via the exchange of decentralized identifiers called DIDs~\cite{DIDs}. DIDs are self-identifiers that enable verifiable digital identities for secure communication. They are designed to be secure and privacy-preserving using cryptographic methods to demonstrate control and ensure trust in the interactions associated with them. 

The DIDComm agent sends messages from the user controller to the communication agent on the service provider side, sharing them with the controller. These agents ensure secure and protected communication of information between users and service providers. Users and service providers will have interactive web interfaces connected to their respective controller components. Similarly, using the DIDComm agents, user controllers can send messages to each other to execute distributed computation. The web interface enables users and service providers to perform all operations efficiently and view responses easily.

\vspace{-2 mm}
\subsubsection{System Functions \\}
\label{functions}

\noindent\textbf{Compute} function enables $SP$ to analyze user data using a privacy-preserving method. It can perform custom functions on the data agent to compute and return derived values without directly exposing personal user data. When a service provider submits a computation request, the $UC$ validates it against access control settings $AC$. Upon approval, the secure computation is executed (within an isolated enclave environment), preventing direct access to raw data. The enclave performs the specified analysis on the authorized subset of data. The framework utilizes compute functionality to run ML models locally and compute particular functions. The outcome of such computations, i.e., data products, can then be used to provide information to the user or the service provider (e.g., for dashboard analytics). An advantage of this approach is that a service provider can compute its functions using a wide variety of data that may be generated by other service providers and stored in users' data agents. When a $SP$ submits a request with permitted operations $OP$, the compute function performs as $Cmp(SP, RE, DS) \rightarrow (r_{Cmp}, \sigma)$, if the operations are among the permitted operations.

\begin{figure}
    \vspace{-5mm}
    \centering
    \includegraphics[width = 0.75\linewidth]{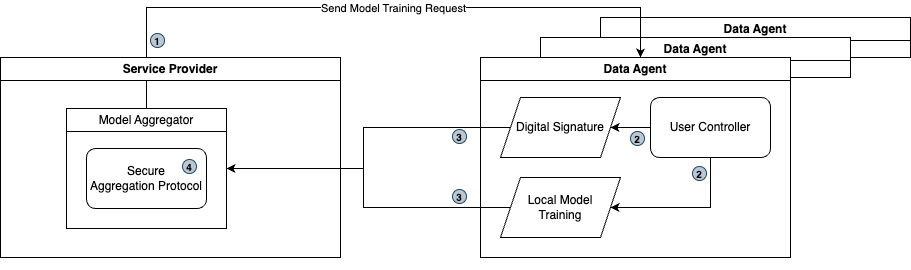}
    \caption{Data flow of the build functionality.}
    \label{fig:build}
    \vspace{-5mm}
\end{figure}

\noindent\textbf{Build} is available to $SP$ (by running their model aggregator) to perform federated learning and train new machine learning (ML) models. Figure~\ref{fig:build} illustrates the data flow of the build functionality. The $UC$ are distributed devices that train models using personal data without directly sharing such data with service providers. The $MG$ and data agent communication is established via DIDComm~\cite{DIDComm}. Hence, the parties can engage in trusted communication without revealing unnecessary personal details of any participating user. 
Upon establishing the communication, the $MG$ and $UC$ must present one or more verifiable data products (e.g., attributes or credentials) to establish trust. Verifiable data products are attributes or credentials that an issuer digitally signs to enable the authenticity and validity of the records.

Once a $UC$ receives a model from the $MG$, verifies the $AC$ settings, trains the model using its data, and sends the trained model's result back to the $SP$. Formally, after initiating the communication between $SP$ and $MG$ and verification of each other's credentials, consequentially, and upon receiving the request, the build function outputs $M^{up}$ as the locally trained model update by aggregating the results running $Build((r_{cmp_1}, \sigma_1),...,(r_{cmp_n}, \sigma_n)) \rightarrow M^{up}$.


\noindent\textbf{Verify} functionality validates i) the correctness (or legitimacy) of user inputs to the functions provided by service providers and ii) the correctness of computation for these functions. It runs cryptographic verifications against the data products of the input user data. Therefore, the system ensures the $SP$'s confidence in the user's data and computation results provided by the $UC$ through secure enclaves. Hence if the signatures on the dataset and computation results are valid, the verify function is satisfied, $Vrf(SP, UC, DS, (r_{cmp}, \sigma)) \rightarrow {1}$.

We leverage the AWS Nitro enclave~\cite{whatIsNitro}, a secure virtual environment that AWS provides for verifiable and trustworthy computations. They allow users to establish isolated computing environments, enhancing the protection and secure processing of sensitive data. It gives service providers trust and confidence in the computation output by the user controller. 

\vspace{-3 mm}
\section{Security and Privacy Analyses}
\label{sec:security}
\vspace{-2 mm}

This Section provides the security theorem of the proposed scheme $\pi$ in Section~\ref{sec:proposed} based on the threat model we provided in Section~\ref{threatmodel}.

We demonstrate user data privacy by showing that service providers learn nothing beyond the computation results and proving that responses to multiple requests from service providers leak no unauthorized information. We define the real and ideal worlds of the data agent and service provider and construct a simulator that can generate views indistinguishable from real protocol executions without access to the actual user data. We show that distinguishing between real and simulated views requires breaking the security of the underlying enclave or the attestation mechanism. The privacy analyses is guarantees under standard computational assumptions and not under information-theoretic (statistical) privacy.







   

\begin{theorem}
For any probabilistic polynomial time (PPT) adversary $\mathcal{A}$ corrupting service provider $SP$, in the proposed architecture $\pi$ in Section~\ref{sec:proposed}, there exists a PPT simulator $\mathcal{S}_1$ such that for all user data $DS$ and computation requests $re_i \in RE$, where $ i= \{1,\ldots,q\}$, the views of a $\mathcal{A}$ are computationally indistinguishable with the following security properties.

\begin{itemize}

\item Privacy: For any datasets $DS_0, DS_1$:
   \[ \{\mathsf{VIEW}_{\mathcal{A}}^{\pi}(DS_0,\{re_i\}_{i=1}^q)\} \approx_c \{\mathsf{VIEW}_{\mathcal{A}}^{\pi}(DS_1,\{re_i\}_{i=1}^q)\}. \]

\item Computation Correctness: For a valid $(r_{cmp}, \sigma)$:
   \[ \Pr[Vrf(SP,UC,DS,(r_{cmp},\sigma)) = 1 \land r_{cmp} \neq Cmp(SP,R,DS)] \leq \mathsf{negl}(k). \]

\item Access Control: For any unauthorized request $R'$:
   \[ \Pr[Allow(RE') = 1] \leq \mathsf{negl}(k) \]

\item Model Integrity: For a valid build request:
   \[ \Pr[Bld(\{(r_{cmp_i}, \sigma_i)\}_{i=1}^q) \neq Agg(\{r_{cmp_i}\}_{i=1}^q)] \leq \mathsf{negl}(k). \]
 
\end{itemize}

\end{theorem}

We analyze the privacy proof comprehensively, and the other protocol properties follow similar arguments. We prove that for any two datasets, the views of a polynomial-time adversary corrupting a service provider are computationally indistinguishable. No polynomial-time adversary can extract additional information about the underlying user data. We guarantee that privacy holds even under multiple executions of the protocol.

 \begin{proof}
 We construct the proof by designing a simulator $\mathcal{S}_1(1^k, re_i, r_{cmp_i})$ for a $re_i \in RE $, where $q$ is sequence of computation requests. The $\mathcal{S}_1$ can generate an indistinguishable view from the real protocol without access to the dataset. The $\mathcal{S}_1$ inputs the $re_i$, the security parameter $1^k$, and a protocol's honest result $\{r_{cmp_i}\} \leftarrow {Cmp}(SP, RE, DS)$ at each request.  Then it generates the simulated signature $\tilde{\sigma}_i$, and adds  $(re_i, \tilde{r_{cmp}}_i, \tilde{\sigma}_i)$ to the simulated view which is $\mathsf{VIEW}_{S_1} = \{re_i, \tilde{\sigma}_i, r_{cmp_i}\}_{i=1}^q$.

 The real world's executed result follows the steps of the protocol, and the real view is $\mathsf{VIEW}_{\mathcal{A}}^\pi = \{re_i, \sigma_i, r_{cmp_i}\}_{i=1}^q$.

 We prove in the following that the simulated view is computationally indistinguishable from the real view of the protocol. 

Assume, there exists a PPT distinguisher $\mathcal{S}_2$ that can differentiate between the simulated protocol view and the real one on dataset $DS$ with non-negligible probability $\epsilon$: $$\left| \Pr[\mathcal{S}_2(\text{VIEW}^{\pi}_{A}(DS, \{re_i\}_{i=1}^q)) = 1] - \Pr[\mathcal{S}_2(S_1(1^k, \{re_i, r_{cmp_i}\}_{i=1}^q = 1] \right| \geq \epsilon$$

We can construct an adversary $\mathcal{B}$ against the enclave security using this distinguisher. $\mathcal{B}$ receives the security parameter $1^k$ and has access to the enclave's oracle $\mathcal{O}$. It has the honest computation results $r_{cmp_i}$. For each $r_{cmp_i}$, the $\mathcal{B}$ queries enclaves oracle $\mathcal{O}(r_{cmp_i)}$. It receives $\theta_i$ that equals to a real output $\sigma_i = Sig(r_{cmp_i})$, or the simulated results $\tilde{\sigma}_i$, depending on the oracle's mode. $\mathcal{B}$ constructs its view as $V = \{re_i, r_{cmp_i}, \theta_i\}_{i=1}^q$. It uses $\mathcal{S}_2$ to run the view and outputs $1$ if and only if $\mathcal{S}_2$'s output is $1$.

 The $\mathcal{B}$ simulates the real view when the oracle provides the real enclave signature. Otherwise, when oracle provides simulated enclaves signature, $\mathcal{B}$ simulates the $\mathcal{S}_1's$ view. Therefore, $\mathcal{B}$'s advantage in distinguishing between the two worlds is non-negligible, and it can break the security of the enclaves. This contradicts the security assumption and completes the proof.

 The proofs for the other defined security properties, including computational correctness, access control, and model integrity, follow similar reduction arguments.
\end{proof}

\vspace{-3 mm}
\section{Evaluation}
\label{sec:evaluation}
\vspace{-2 mm}

This section presents a practical implementation and evaluation of our privacy-preserving architecture. Through a functional prototype, we assess the proposed architecture's feasibility throughout the complete lifecycle of collecting, storing, and processing individuals' sensitive information while maintaining privacy guarantees.

To demonstrate the feasibility of our architecture, we have developed and deployed \footnote{All implementation codes are available on GitHub and will be provided upon request.} a prototype incorporating the core components described in Section~\ref{sec:architecture}. The details of implementation, deployment, and testing use cases are as follows.

The data plug component is the primary mechanism for securely retrieving information from third-party service providers. We implement specialized data plugs for multiple platforms, including: (i) a Reddit API integration that collects social media activity data (posts, likes, dislikes), (ii) a Spotify API connection that retrieves user profile and music preference data and (iii) a direct upload functionality compatible with Google Takeout exports. The modular approach allows users to connect their accounts, provide necessary authentication credentials, and designate specific service providers as authorized data sources. Data plug implementation can be readily extended to incorporate diverse source APIs for platforms such as  Facebook, LinkedIn, Google Maps, or any educational information systems (as discussed in our application scenarios in Section~\ref{sec:application}). It makes the proposed system compatible with the most popular service providers for its use in various application areas.

We have developed a web-based interface for comprehensive access control settings using AWS Attribute-Based Access Control~\cite{ABAC} as the underlying mechanism. This implementation gives users fine-grained permission control over their data and its usage contexts, ensuring that service providers can only access information explicitly authorized by the users. 

The user controller component forms the operational core of our implementation. All incoming requests from service providers are processed through an auditing system that verifies permissions against the established access control settings. Only requests with valid permission from the user are processed and executed, and the results are shared. The system supports both computation requests and model training operations across multiple data sources. 

We evaluated our system's computation capabilities using data collected from the Reddit API (saved, liked, and disliked posts). Using data collected from the Reddit API, we apply a Natural Language Processing (NLP) technique called Name Entity Recognition (NER) to identify music artists mentioned in posts data collected from the Reddit platform. These extracted preferences were then shared with a music service provider (Spotify) to enhance recommendation relevance without exposing the user's raw data. Secondly, we implemented sentiment analysis computation on YouTube interaction data, calculating average sentiment scores across user comments on watched videos. This provided service providers with valuable engagement metrics while preserving user privacy. 

Our implementation allows service providers to select from various computational approaches, including linear regression, statistical aggregations, and custom functions. They can also specify which user data portions should be included in the analysis, enhancing flexibility while maintaining privacy boundaries.

\begin{figure}
    \centering
    \includegraphics[width = 0.5\linewidth]{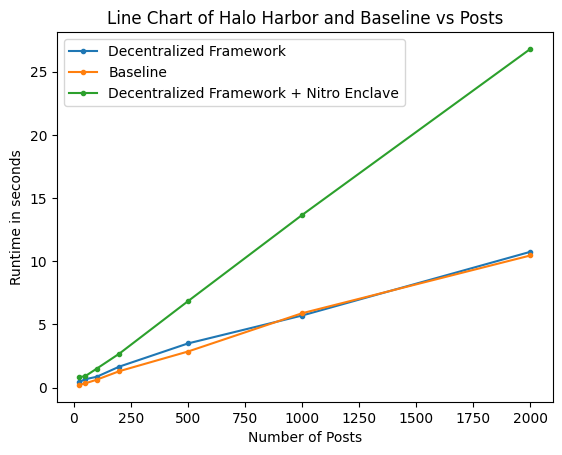}
    \caption{Runtime of name entity recognition on Reddit data: centralized vs. decentralized.}
    \label{fig:compRuntime}
    \vspace{-5mm}
\end{figure}

To evaluate the framework's usability, we conducted a comparative analysis of the computation runtime. Specifically, we compare the runtime of computation in a centralized setting (baseline) with that observed in our decentralized framework (considering scenarios both with and without the utilization of a secure enclave). For evaluation, the system is deployed locally on a Windows machine equipped with an Intel(R) Core(TM) i7-10750H CPU and 16 GB of RAM. For a secure enclave, we deploy an AWS Nitro Enclave inside Amazon Elastic Container Service, utilizing 2 m5xlarge CPUs and 4 GB of RAM.  We also assess systems' scalability by incrementally increasing the data size. Figure \ref{fig:compRuntime} shows that the runtime rises linearly with the volume of processed data (number of posts), with our system demonstrating comparable performance to the centralized baseline despite its enhanced privacy protections. The runtime for computations executed within the Nitro Enclave does exceed that of the decentralized framework. This disparity can be attributed to the inherent overhead associated with loading and executing computations within the isolated environment of the secure enclave, coupled with the differential in computational resources at the time of testing. The Nitro Enclave, while providing enhanced privacy guarantees, comes with a tradeoff between efficiency and elevated security features (privacy of the $SP$'s computation functions and verifiability of the results). As the Nitro Enclave is hosted on AWS Elastic Compute Cloud (EC2), its computational capacity (CPU and memory allocation) can be increased to potentially mitigate runtime and enhance overall efficiency. This indicates that our privacy-preserving architecture maintains efficiency without significant computational overhead.

Similarly, we successfully implemented and tested a neural network model trained on data from multiple users using federated learning. Using data collected from the Reddit API,  a classification model was trained to predict user preference, specifically whether a given title would elicit positive engagement (i.e., "liked") from users. Each user's data agent receives and executes the training request independently according to the permission of its access control setting. The data agents send their respective model weight updates to the service provider agent via secure DIDComm messages. 

On the service provider side, the "service provider controller" component aggregates these weights and updates the model (detailed further in Section~\ref{seviceprovideragent}), then coordinates subsequent training rounds. This implementation offers service providers considerable flexibility in specifying data features for training, selecting from standard machine learning models or implementing custom architectures, and configuring various hyperparameters such as the number of training rounds. 

Our implementation demonstrates that robust privacy protection and valuable data utilization can coexist within a properly designed architecture, addressing the fundamental limitations of existing approaches.


\vspace{-3 mm}
\section{Applications}\label{sec:application}
\vspace{-2 mm}

Our proposed decentralized privacy-preserving architecture finds multifaceted applications across several industries. In healthcare, technology can aid in pandemic prevention and monitoring by leveraging wearable devices and self-reported formats. Workforce development and enhancement can benefit through personalized data collection relating to an employee's work skills and well-being metrics, ensuring a healthier and more efficient workforce. Students can centralize all their academic and skill-related achievements within education and training, making it easier for prospective employers or educational institutions to evaluate their capabilities. In transportation, our architecture can help rental companies provide a seamless, secure, and customized experience for their customers by storing essential driving-related data.

\noindent\textbf{Healthcare} data collected from the wearable devices and apps is stored in the personal data agents rather than shared directly with service providers. Collected data is then used for i) executing local AI models to provide insights to the users; ii) developing new AI models via federated learning; and iii) privacy-preserving data sharing with the medical providers.  

Similarly, users can aggregate data from fitness trackers, such as Google Fitbit, which monitor heart rate, sleep patterns, and physical activity levels, along with data from health apps, nutrition tracking apps, and other related applications. Healthcare providers can access insights derived from this data while users maintain privacy and control, enabling personalized treatment recommendations without exposing raw data.

Another use case applies to collecting employee health data to promote workforce efficiency and wellness. This system: i) aggregates stress-related metrics from various sources and provides personalized recommendations to employees; ii) develops improved stress-level inference models and facilitates team-building through privacy-preserving compatibility assessments. 

Furthermore, healthcare professionals can connect productivity tracking applications to generate a comprehensive profile when seeking employment opportunities. Employers can receive cryptographic proofs of the employee's qualifications and skills, ensuring authenticity while respecting privacy.

\noindent\textbf{Education and training} includes transcript and skill management for students. Collection of all skills a student has gained (including transcripts, training, reference letters/endorsements, books, or videos) in the student's user agent. Processing such material to extract the skills and providing a set of particular skills (in a verifiable way) to a potential future employer or school is an application of our proposed scheme. The system can also provide privacy-preserving dashboarding. It can also create AI models using the data stored in students' user agents.

\noindent\textbf{Usage-Based Auto Insurance (UBI)} determines auto insurance premiums based on individual driving behavior, mileage, and data collected from vehicles through a telematics device. This device records crucial metrics like speed and braking patterns. First, it sends the data to the driver's Data Agent, which processes it locally to generate a driving score without revealing raw data.

The Data Agent employs verifiable computation to protect privacy, creating a cryptographic signature for the driving score that reflects the driver's behavior. Insurance providers can assess risk and calculate premiums based on this score while maintaining data confidentiality and privacy. Drivers can selectively disclose their driving data, and as the Data Agent updates the score over time, it can adjust premiums accordingly.

This integration enhances privacy, enables accurate information sharing, and builds trust between drivers and insurers. It also allows storing driving credentials in a digital wallet linked to the driver's decentralized identity, facilitating potential sharing with other insurers and promoting fairness in premium calculation.

\noindent\textbf{Personalized Treatment Planning with a Digital Twin} replicas of physical entities like organs, systems, or entire patients are increasingly being used to simulate and analyze various medical conditions and treatment options. Data agents are crucial in managing and securing vast amounts of sensitive data in creating and utilizing digital twins. For example, consider a patient named Bob undergoing treatment for a complex cardiac condition. His healthcare provider creates a digital twin of his heart. A detailed virtual model that simulates how his heart functions under different conditions and treatment scenarios. Cardiologists use this digital twin to plan and optimize Bob's treatment. 

Throughout this process, data agents are essential in securely managing the collection and integration of data from various sources, including Bob's medical records, imaging data, genetic information, and real-time data from wearable devices. The data agent ensures that all of Bob's sensitive health data is decentralized, giving him complete control over who can access it. During the simulation phase, the data agent facilitates the secure sharing of necessary data with simulation tools, ensuring that only authorized parties can contribute to the analysis without directly accessing Bob's raw data. 

\noindent\textbf{Sports Field Fan Engagement} employs Decentralized Identifiers (DID) to verify credentials, thereby enhancing security, compliance, and fan engagement. The system stores digital credentials on a decentralized network, which enables tamper-proof validation at venue entry points. The system facilitates age verification at point-of-sale terminals without compromising personal information. Furthermore, this credential infrastructure supports an integrated loyalty program, allowing fans to redeem their credentials for various rewards. The system also has potential applications for regulatory compliance in sports betting platforms.

 \noindent\textbf{Government Services and Administration}:
Our system can enhance government operations by improving service delivery while protecting citizen privacy. It allows eligibility verification for social benefits without disclosing detailed income, enables secure inter-agency information sharing, and facilitates anonymous census data collection. Companies can prove contract compliance without revealing proprietary information, and a digital identity system can ensure secure e-government access. Additionally, it supports verifiable participation in public consultations and allows travelers to prove visa compliance without sharing travel history.

\vspace{-3 mm}
\section{Discussion}
\label{sec:discussion}
\vspace{-2 mm}

This section provides our proposed framework's functionality and security extension. It includes a data-sharing functionality that enables users to share their data and data products in a privacy-preserving way with service providers and other users. Moreover, the system can support more robust security mechanisms, as discussed here.

\vspace{-2 mm}
\subsection{Extend Framework-Share Functionality}
\vspace{-2 mm}

One promising direction involves defining a "share" functionality and implementing fine-grained data-sharing policies through attribute-based encryption (ABE). Access to shared data will be governed by complex policies based on the attributes of both the data and the requesting entities. It allows for more nuanced control over data sharing, where users can specify complex conditions under which their data can be accessed and used while ensuring that these policies are cryptographically enforced rather than relying solely on access control lists. 

 We focus on sharing individual data under differential privacy~\cite{dwork2006differential}. We demonstrated~\cite{emre_codaspy} how the original definition of local differential privacy to preserve the privacy of individuals in data sharing with an untrusted data collector is unsuitable for sharing personal data streams that include correlations. We will integrate the differentially private data-sharing algorithms developed on the user controllers through the computing functionality. For differentially-private user data sharing, we will allow service providers to specify what type of data they wish to receive and the nature of operation they are willing to do using the received data. Using this information and the consent of the users, we will use the algorithm to obfuscate the users' data (by also considering the correlations within the data) and apply post-processing steps (considering the nature of the target operation) to maximize the utility of the shared data. 

In addition to these core functionalities, our architecture can incorporate Zero-Knowledge Proofs (ZKPs) as a fundamental privacy-preserving technology. These protocols, such as Bulletproofs \cite{bunz2018bulletproofs}, Groth's SNARKs \cite{groth2016size}, and zk-SNARKs \cite{ben2014succinct}, enable data agents to prove the validity of statements about user data without revealing the data itself. Hence, they allow secure verification of user credentials, attributes, and computations while maintaining user privacy. Deploying ZKP enhances the ability of data agents to interact with service providers in a privacy-preserving manner, further distinguishing our decentralized architecture from traditional centralized solutions. 


\vspace{-2 mm}
\subsection{Extended Security}
\vspace{-2 mm}

We proved the proposed framework's security against semi-honest data providers. The security model can be extended to address more genteel attack vectors and provide more robust privacy guarantees.  

One potential extension involves incorporating differential privacy mechanisms into the computation phase. The noise should be calibrated and added to the computation results before they leave the secure enclave. This approach would provide formal privacy guarantees against inference attacks while maintaining utility for legitimate computations. 

Additionally, the system can be extended to support secure multi-party computation (SMPC) protocols, allowing multiple service providers to jointly compute functions on encrypted data without revealing their inputs. It enables more complex cross-organizational data analyses while preserving privacy.

The security framework could also be strengthened by implementing advanced zero-knowledge proofs demonstrating the correctness of computations without revealing intermediate values or implementation details. This enhancement forces the parties to follow the protocol instructions and execute the functions correctly. It enables service providers to verify the computations within a secure enclave to check that specific algorithms were executed correctly, thereby increasing transparency and trust in the system while maintaining the confidentiality of proprietary computation methods.

\vspace{-3 mm}
\section{Conclusion}
\label{sec:conclusion}
\vspace{-2 mm}

We introduced a groundbreaking architecture that fundamentally re-imagined personal data management, enabling users to control their sensitive information while allowing service providers to derive valuable insights through privacy-preserving computations. We proposed a decentralized, privacy-preserving architecture that addresses the inherent shortcomings of traditional centralized systems regarding privacy and security. The proposed framework mitigates data abuse or loss of privacy by allowing users to retain complete control and ownership over their sensitive data. By integrating advanced privacy-enhancing technologies, such as secure enclaves, verifiable computation, and federated learning, the framework allows service providers to execute secure and verifiable computations and model training while ensuring that sensitive information remains protected. This architecture offers significant advantages regarding user privacy, data security, and transparency, providing a scalable and trustworthy solution for the future of data-driven services across sectors such as education, healthcare, and finance. 

In future work, we will extend the framework to support more complex interactions between data agents, enabling broad collaborative computations while maintaining strong privacy guarantees. The architecture will incorporate zero-knowledge proofs, secure multi-party computations, or differential privacy to support data analysis and sharing capabilities across domains, ensuring long-term sustainability and widespread adoption of privacy-preserving data management solutions.


%
%
%
%
\newpage
\bibliographystyle{splncs04}
\bibliography{sample}

\end{document}